\documentstyle[psfig]{mn}

\def\ga{\mathrel{\hbox{\rlap{\hbox{\lower4pt\hbox{$\sim$}}}\hbox{$>$}}}}
\def\fd{\hbox{$.\!\!^{\rm d}$}}

\title[High-speed photometry of faint Cataclysmic Variables: III.]
{High speed photometry of faint Cataclysmic Variables: III. V842 Cen,
BY Cir, DD Cir, TV Crv, V655 CrA, CP Cru, V794 Oph, V992 Sco, EU Sct 
and V373 Sct}
\author[Patrick A. Woudt and Brian Warner]
       {Patrick A. Woudt\thanks{E-mail: pwoudt@circinus.ast.uct.ac.za} 
        and Brian Warner\thanks{E-mail: warner@physci.uct.ac.za}\\
        Department of Astronomy, University of Cape Town, Private Bag,
        Rondebosch 7700, South Africa}
\date{}

\begin{document}

\maketitle

\begin{abstract}
We present further results from a high speed photometric survey
of faint cataclysmic variables. We find that V842 Cen (Nova Cen 1986 No.~2) is 
highly active, but with no evident orbital modulation. BY Cir (Nova Cir 1995) 
is an eclipsing system with an orbital period ($P_{orb}$) of 6.76 h. TV Crv,
an SU UMa type dwarf nova, is found to have $P_{orb}$ = 1.509 h from
photometry at quiescence. DD Cir (Nova Cir 1999) is an eclipsing system
with $P_{orb}$ = 2.339 h, a possible secondary eclipse, and a $\sim$670 s photometric
modulation. V655 CrA is highly active but shows no orbital modulation.
The identification of V794 Oph is probably incorrect as we find no photometric variability.
CP Cru (Nova Cru 1996) is an eclipsing system with $P_{orb}$ 
= 22.7 h. V992 Sco has $P_{orb}$ = 3.686 h from its periodic modulation.
The supposed identification of EU Sct is probably incorrect. And finally,
during one run V373 Sct (Nova Sct 1975) had a 258.3 s coherent periodicity,
making it a candidate DQ Herculis star.
\end{abstract}


\section{Introduction}

             In continuation of this series (Woudt \& Warner 2001,2002) we
present the results of high speed photometry of a further 10 cataclysmic
variable stars (CVs). These are mostly quite faint and in crowded fields,
requiring the use of the University of Cape Town's CCD photometer
(O'Donoghue 1995) in frame transfer mode and with no optical filter (i.e.
the observations were made in `white light'). All of our observations were
made at the Sutherland site of the South African Astronomical Observatory,
using the 1.9-m (74-in) and 1.0-m (40-in) reflectors. The photometry was
calibrated by observing hot white dwarf standards, but is not considered of
high quality because CVs have non-standard flux distributions, so
transformations to a standard photometric system (and, indeed, choice of the
correct atmospheric extinction coefficient) are not possible. We leave our
magnitudes on our own non-standard system.

    As in the previous papers, we have concentrated on faint nova remnants.
These have proved to be rewarding (e.g. for discovery of magnetic
associated phenomena) and are in any case in most need of increased
knowledge of orbital periods and of the discovery of eclipsing systems. Our
observational philosophy is to consider this work as a survey, spending only
sufficient time on each star to establish its essential properties. For the
faintest objects where no certain identification has hitherto been given, we
are usually satisfied merely to discover which of the candidates in the
field shows rapid brightness variations. Often such a demonstration requires
several attempts before the seeing conditions are good enough to isolate
candidates from their neighbours. We use psf fitting, aperture photometry
and image subtraction techniques in this quest. In the survey we are
frequently pushing the telescopes to their magnitude limits and adjust the
integration time according to seeing or faintness. We only observe brighter
objects if the seeing is poor, or there is intermittent cloud, or there is a
proximate moon.

In Section 2 we give the results of our observations; Table 1 contains
the list of runs and the details for all the observed stars. 
Section 3 contains a summary of the results obtained from these new observations.

\begin{table*}
 \centering
% \begin{minipage}{160mm}
  \caption{Observing log.}
  \begin{tabular}{@{}llrrrrrcc@{}}
 Object       & Type         & Run No.  & Date of obs.          & HJD of first obs. & Length    & $t_{in}$ & Tel. &  V \\
              &              &          & (start of night)      &  (+2451000.0)     & (h)       &     (s)   &      & (mag) \\[10pt]
%{\bf Nova(?) Car 1971}& NR   & S6293    & 2002 Mar 12 &             &             &           &   & 18.0: \\[5pt]
{\bf V842 Cen}& NR           & S6096    & 2000 Jun 03 &  699.24431  &  10.23      &       5   &  74-in & 15.8\\
              &              & S6097    & 2000 Jun 04 &  700.20473  &  11.14      &   5, 10   &  74-in & 15.8\\
              &              & S6321    & 2002 Mar 21 & 1355.33839  &   1.68      &       6   &  40-in & 15.9\\[5pt]
{\bf BY Cir}  & NR           & S6298    & 2002 Mar 13 & 1347.56636  &   1.93      &      15   &  74-in & 17.8$^*$\\
              &              & S6301    & 2002 Mar 14 & 1348.43030  &   5.29      &      15   &  74-in & 17.8$^*$\\
              &              & S6302    & 2002 Mar 15 & 1349.26660  &   9.15      &      20   &  74-in & 17.7$^*$\\
              &              & S6306    & 2002 Mar 18 & 1352.48120  &   1.25      &      15   &  74-in & 17.7$^*$\\
              &              & S6308    & 2002 Mar 18 & 1352.60138  &   1.39      &      20   &  74-in & 17.7$^*$\\[5pt]
{\bf DD Cir}  & NR           & S6399    & 2002 May 31 & 1426.21021  &   2.99      &      60   &  74-in & 20.1$^*$\\     
              &              & S6404    & 2002 Jun 02 & 1428.22039  &   5.83      &      60   &  74-in & 20.1$^*$\\     
              &              & S6408    & 2002 Jun 03 & 1429.21094  &   1.02      &      60   &  74-in & 20.2$^*$\\     
              &              & S6416    & 2002 Jun 05 & 1431.20365  &   2.38      &      60   &  74-in & 20.2$^*$\\     
              &              & S6421    & 2002 Jun 08 & 1434.20882  &   3.01      & 60, 120   &  74-in & 20.1$^*$\\     
              &              & S6425    & 2002 Jun 09 & 1435.19953  &   4.87      &      90   &  74-in & 20.0$^*$\\[5pt]     
{\bf TV Crv}  & DN           & S6355    & 2002 Apr 07 & 1372.32645  &   2.88      &      30   &  74-in & 18.8\\
              &              & S6359    & 2002 Apr 08 & 1373.23083  &   1.58      &      30   &  74-in & 18.9\\[5pt]
{\bf V655 CrA}& NR           & S6505    & 2002 Aug 28 & 1515.33190  &   2.71      &      10   &  74-in & 17.6\\
              &              & S6512    & 2002 Aug 29 & 1516.25599  &   1.45      &      15   &  74-in & 17.4\\
              &              & S6517    & 2002 Aug 31 & 1518.22488  &   5.59      &      15   &  74-in & 17.6\\
              &              & S6521    & 2002 Sep 01 & 1519.21711  &   3.21      &      15   &  74-in & 17.6\\[5pt]
{\bf CP Cru}  & NR           & S6294    & 2002 Mar 12 & 1346.32945  &   7.59      &  45, 60   &  74-in & 19.6$^*$\\     
              &              & S6303    & 2002 Mar 16 & 1350.30020  &   8.08      &      45   &  74-in & 19.7$^*$\\     
              &              & S6304    & 2002 Mar 18 & 1352.24923  &   2.02      &      45   &  74-in & 19.7$^*$\\     
              &              & S6314    & 2002 Mar 19 & 1353.45168  &   1.61      &      60   &  40-in & 19.6$^*$\\     
              &              & S6325    & 2002 Mar 22 & 1356.47462  &   1.53      &      60   &  40-in & 19.7$^*$\\     
              &              & S6332    & 2002 Mar 23 & 1357.46866  &   0.88      &      60   &  40-in & 19.6$^*$\\[5pt]
{\bf V794 Oph}& NR           & S6419    & 2002 Jun 06 & 1342.33235  &   2.06      &       6   &  74-in & 17.2:\\     
              &              & S6429    & 2002 Jun 10 & 1346.29756  &   1.26      &      15   &  74-in & 17.1:\\[5pt]
{\bf V992 Sco}& NR           & S6322    & 2002 Mar 21 & 1355.41396  &   5.76      &      20   &  40-in & 17.2\\
              &              & S6326    & 2002 Mar 22 & 1356.54380  &   2.69      &      10   &  40-in & 17.0\\
              &              & S6339    & 2002 Mar 24 & 1358.49867  &   3.51      &      15   &  40-in & 17.0\\
              &              & S6398    & 2002 May 30 & 1425.28958  &   9.71      &       6   &  74-in & 17.1\\
              &              & S6400    & 2002 May 31 & 1426.34201  &   2.81      &      10   &  74-in & 17.0\\
              &              & S6402    & 2002 Jun 01 & 1427.28741  &   1.08      &      10   &  74-in & 17.1\\
              &              & S6405    & 2002 Jun 02 & 1428.47032  &   3.16      &      10   &  74-in & 17.1\\[5pt]
{\bf EU Sct}  & NR           & S6500    & 2002 Aug 27 & 1514.44263  &   0.62      &       5   &  74-in & -\\     
              &              & S6523    & 2002 Sep 01 & 1519.40923  &   0.83      &   5, 15   &  74-in & -\\[5pt]
{\bf V373 Sct}& NR           & S6358    & 2002 Apr 07 & 1372.57518  &   2.28      &      30   &  74-in & 18.7\\     
              &              & S6361    & 2002 Apr 08 & 1373.48768  &   4.31      &      30   &  74-in & 18.6\\     
              &              & S6497    & 2002 Aug 27 & 1514.24270  &   2.90      &      30   &  74-in & 18.5\\[5pt]
\end{tabular}
{\footnotesize 
\newline 
Notes: NR = Nova Remnant; DN = Dwarf Nova; $t_{in}$ is the integration time; `:' denotes an uncertain value; $^*$ mean magnitude out of
eclipse.\hfill}
\label{tab1}
%\end{minipage}
\end{table*}

\section{Observations}

%\subsection{Nova (?) Carinae 1971}
%
%Downes \& Duerbeck (2000) point out that the supposed Nova Cir
%1971 was not detected on a direct photograph of the sky taken 22 days after
%the discovery spectrum, so the original detection was probably an emulsion
%defect. We observed (Table 1) the candidate star indicated on Duerbeck's
%(1987) finding chart and found no variability.

\subsection{V842 Centauri}

\begin{figure*}
\centerline{\hbox{\psfig{figure=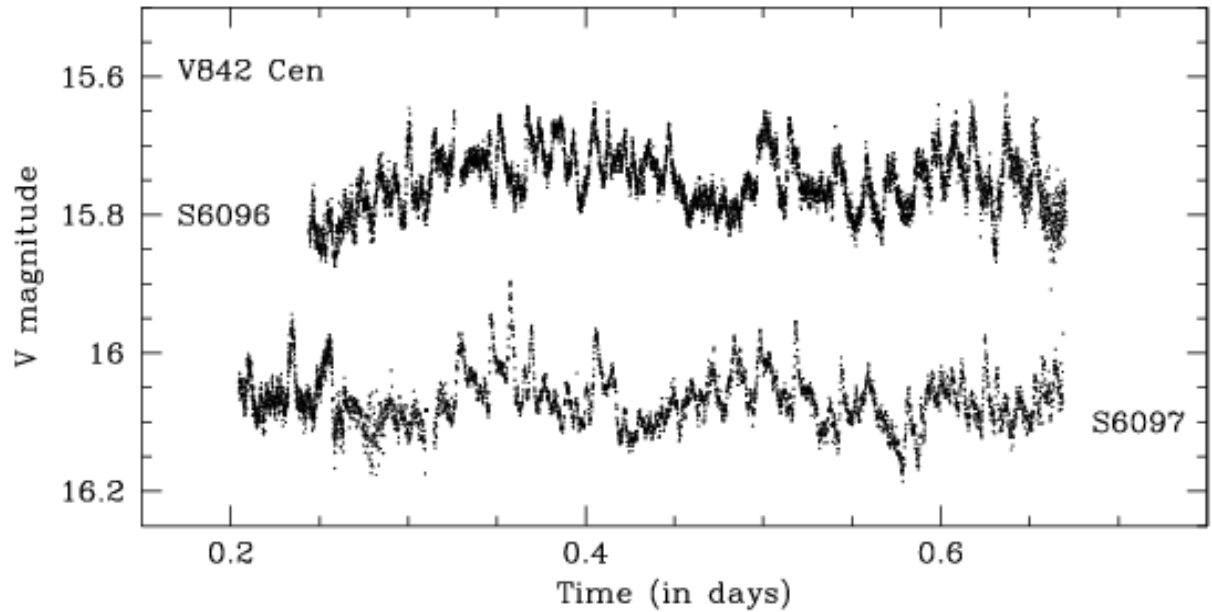,width=16.0cm}}}
  \caption{The light curves of V842 Cen, obtained in June 2000. The light curve of run
S6097 has been displaced vertically downwards by 0.3 mag for display purposes only.}
 \label{lcv842cen}
\end{figure*}

\begin{figure*}
\centerline{\hbox{\psfig{figure=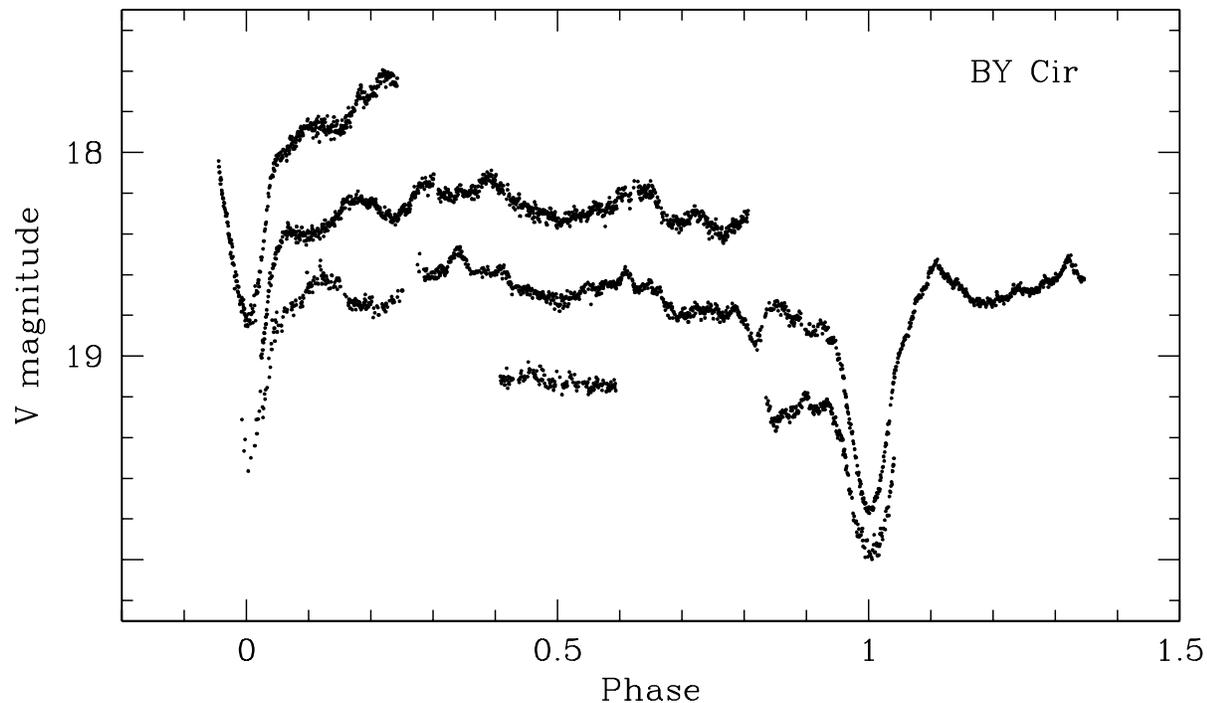,width=16.0cm}}}
  \caption{The light curves of BY Cir, obtained in March 2002. All light curves, except the
upper one, have been displaced downwards by 0.5, 1.0, and 1.5 mag, respectively, for display
purposes. The light curves have been phased according to emphemeris given in Eq.~\ref{ephbycir}.}
 \label{lcbycir}
\end{figure*}
   
    V842 Cen was Nova Centauri 1986 No. 2, discovered at magnitude 5.6
but subsequently assigned a maximum magnitude of V = 4.6 (Duerbeck
1987). It was a moderately fast nova, with $t_3$ = 48 d, which developed an
obscuring dust shell. De Freitas Pacheco et al. (1989) determined E(B-V) =
0.54 mag, and Downes \& Duerbeck (2000) measured V = 15.82 in 1998
and detected a $6''$ diameter ejecta shell around the nova remnant. An IUE
spectrum has been published (Gonzalez-Riestra, Orio \& Gallagher 1998),
obtained in 1991, showing the strong emission lines of CII, SiIV, NIV, HeII
and CIII characteristic of a post nova with a white dwarf primary that is
still very hot.
   
We have made two long runs on V842 Cen, listed in Table 1 and shown in
Fig.~\ref{lcv842cen}. It is evident that V842 Cen is continuously flaring with rises of
up to 0.25 mag on a characteristic time scale of $\sim$5 mins. This is one of
the most active nova remnants we have seen. Although the light curves are both
$\sim$10.5 h long, we see no clear indication of any orbital modulation and conclude
that V842 Cen is probably of low inclination. We have made Fourier
transforms (FTs) of the runs, and subsections of them, and find that there
are no short period (tens of seconds) modulations that would be classified as
dwarf nova oscillations (DNOs). However, occasionally there appear to be strong 
quasi-periodic oscillations (QPOs) present with
time scales $\sim$750 s and $\sim$1300 s on the first night and $\sim$950 s on the
second night. These QPOs will be analysed and discussed elsewhere, together with results
from other CVs. We point out, however, the striking similarity of the light curve
of V842 Cen to that of the nova-like TT Ari (see Figure 9 of Patterson 1994).

\subsection{BY Circini}

    BY Cir was Nova Circini 1995, discovered at V $\sim$ 7.2 on 27 January 1995
(Liller 1995) and reported to be at V = 15.9 on 23 March 1998 (Downes \&
Duerbeck 2000). The light curve is illustrated by Bateson \& McIntosh
(1998); it shows that BY Cir was a slow nova, falling smoothly with $t_3$ = 157
d.
    
We observed BY Cir on four nights (Table 1), finding on the first night
that it is an eclipsing system with eclipses $\sim$0.9 mag deep. The following
two nights we made longer runs to determine the orbital period, and
subsequently another short run to improve the accuracy of the period. BY
Cir had faded to $\sim$17.8 in the 4 years since Downes and Duerbeck observed
it. The light curves are displayed in Fig.~\ref{lcbycir}, phased at the orbital
period of 6.76 h. 
This is quite long for a nova: BY Cir joins BT Mon ($P_{orb}$ = 8.01 h) and
QZ Aur ($P_{orb}$ = 8.58 h)
as a long period deeply eclipsing nova remnant. The ephemeris for the times
of mid-eclipse is

\begin{equation}
  {\rm HJD_{min}} = 2452347.5790 +  0{\fd}2816 (\pm 2) \, {\rm E}.
 \label{ephbycir}
\end{equation}

The depth of eclipse is seen to vary by about 0.3 mag and there is variable
asymmetry in the eclipse profile. Flickering is present on a number of time
scales, and the profile of the final eclipse has clusters of points showing
that the flickering continues during at least the falling and rising portions.
   
An FT of the combined light curves (with the eclipses removed) shows no
evidence for any persistent coherent periodicities other than $P_{orb}$ and its
harmonics; on individual nights there are no dwarf nova oscillations (DNOs:
periodicities $\sim$10--50 s) detected.

\subsection{DD Circinus}

\begin{figure}
\centerline{\hbox{\psfig{figure=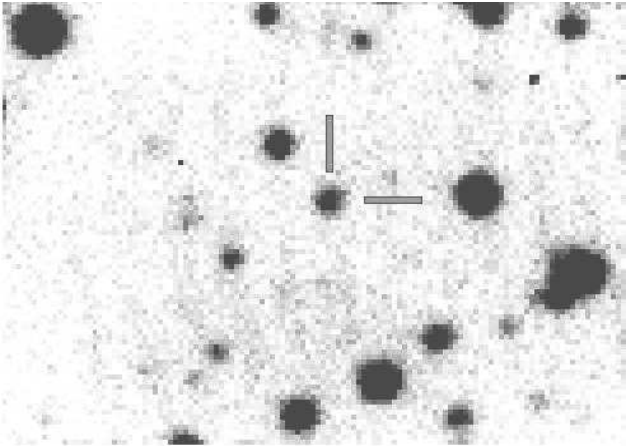,width=8.3cm}}}
  \caption{The finding chart of DD Cir. DD Cir is marked with bars. The field of view
is 50 by 34 arcsec, north is up and east is to the left.}
 \label{fcddcir}
\end{figure}

    DD Cir was discovered as Nova Cir on 23 August 1999 at an apparent
magnitude of 7.7 (Liller 1999). It proved to be a very fast nova, with a 
pre-eruption brightness V $>$ 21 (Downes et al. 2001) and 
$t_2$ $\sim$ 4.5 d. There are no
recent estimates of its brightness. The chart published by Downes et al. does
not identify DD Cir, so we reproduce one of our CCD frames in 
Fig.~\ref{fcddcir}.
    
Our photometric observations are listed in Table 1 and show that DD Cir
has a mean magnitude of $\sim$ 20.1. Maximum brightness during eruption may
have been missed, and it is not certain that it has yet reached its 
quiescent
magnitude, so we can only deduce that the range is $\ge$ 12.5 mag. With the
inclination $\sim 79^{\circ}$ deduced below, DD Cir should have a 
range $\sim$ 15 mag (see Figure 5.4 of Warner 1995) and
therefore may fade out of sight in the next few years.
   
\begin{figure*}
\centerline{\hbox{\psfig{figure=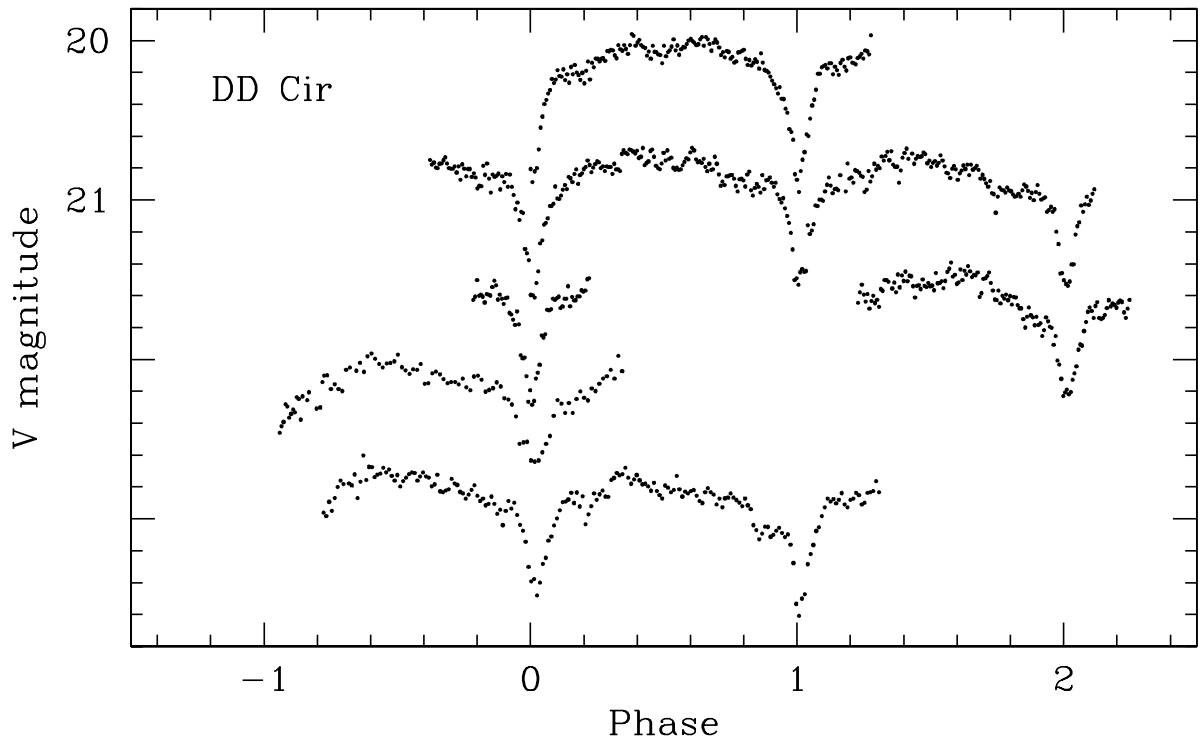,width=16.0cm}}}
  \caption{The light curves of DD Cir. All light curves, except the
upper one, have been displaced vertically downwards by 0.7, 1.4, 2.1, and 2.8 mag, respectively, for display
purposes. The light curves have been phased according to emphemeris given in Eq.~\ref{ephddcir}.}
 \label{lcddcir}
\end{figure*}

The light curves of DD Cir are displayed in Fig.~\ref{lcddcir} and show 
that it is an eclipsing system with an eclipse depth $\sim$ 0.6 mag and 
an orbital period of 2.339 h. A period of that value is of particular 
significance because it lies
within the `period gap' shown by dwarf novae and nova-like variables
(Warner 1995), further reducing the case for such an orbital period gap in
the classical novae (e.g. Warner 2002). The ephemeris for time of mid-eclipse, 
deduced from our light curves, is

\begin{equation}
  {\rm HJD_{min}} = 2452426.2102 +  0{\fd}09746 (\pm 1) \, {\rm E}.
 \label{ephddcir}
\end{equation}

    The upward convexity and variability of the light curve between eclipses
encouraged us to look for a superhump signal, but removing the eclipses
from the light curves, or prewhitening at the orbital period and its
harmonics, leaves no indication of any signal near to $P_{orb}$ in the FTs.
Therefore we interpret the convexity as a reflection effect, resulting from 
the fact that DD Cir was a nova less than three years ago and the primary may
still be quite hot. The high inclination favours such a reflection effect.
     
\begin{figure}
\centerline{\hbox{\psfig{figure=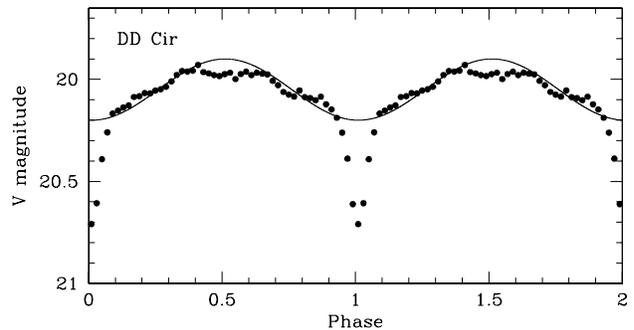,width=8.3cm}}}
  \caption{The average light curve of DD Cir (excluding run S6425), folded on
the orbital period of 2.339 h. A sinusoidal fit to the
light curves (with eclipses removed) is overlayed on the average light curve.}
 \label{lcddcirav}
\end{figure}

A sinusoidal fit to the mean light curve outside of eclipses gives a 
peak-to-peak range of 0.30 mag for the reflection signal -- 
see Fig.~\ref{lcddcirav}. (For
this mean light curve we have omitted the final run, S6425, which shows
deep dips and asymmetry.) Reflection effects in post-novae are expected to
be large only when a high mass transfer rate (high $\dot{M}$), high luminosity
accretion disc does not dominate the system. A relatively short period and
high inclination system such as DD Cir is optimal for a strong reflection
effect. Alternatively, if the system is strongly magnetic (as in a polar) 
then
no disc exists and reflection luminosity can also dominate, as in the nova
V1500 Cyg (Kaluzny \& Chlebowski 1988). In DD Cir, however, there is
clear evidence of an accretion disc: the eclipse profile is that of partial
eclipse of an extended disc.

This can be made more quantitative as follows. The total width of eclipse
is 2$\phi_{d}$ = 0.19 and the total width at half depth is 2$\phi_{p}$ 
= 0.074. Inserted into Equation (2.65) of Warner (1995) 
(which is correct for deep eclipses: because part of the reflected light
remains at mid-eclipse, the eclipse is actually a little deeper than measured)
this gives a disc radius $r_{d}$/$a$ $\ge$ 0.47, 
where $a$ is the separation of components. At $P_{orb}$ = 2.34 h we expect 
the mass of the secondary to be 0.19 M$_{\odot}$ (Equation (2.100) of 
Warner 1995), and adopting a primary mass of 1.1 M$_{\odot}$ 
(in accordance with the typically high masses for
the primaries of novae) we have a mass ratio q = 0.17. Harrop-Allin and
Warner (1996) find from analysis of many CV eclipses that high $\dot{M}$ discs
have radii comparable to the tidal truncation radius $r_{t}$/$a$ = 
$0.60 (1+q)$, which is $r_t/a$ = 0.51 for $q$ = 0.17. It is clear, 
therefore, that DD Cir contains a high $\dot{M}$ disc.
    
The individual light curves, and the mean curve (Fig.~\ref{lcddcirav}), 
show that
the brightness at phase 0.5 falls below what would be expected of a
sinusoidal reflection effect. It is possible that this is the effect of a 
secondary
eclipse. That such might be detectable in special circumstances is shown by
the following analysis.
    
\begin{figure}
\centerline{\hbox{\psfig{figure=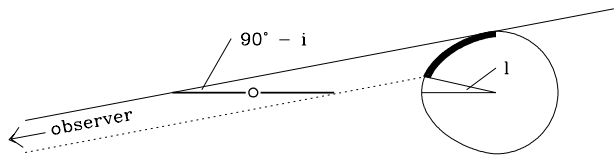,width=8.3cm}}}
  \caption{A schematic representation (not to scale) of DD Cir. }
 \label{ddcirdr}
\end{figure}

With the parameters for DD Cir deduced above, and using the graphs of
the relationships between $q$, inclination $i$, and $\phi_{p}$ given 
by Horne (1985), we find $i = 79^{\circ}$ for DD Cir. Ignoring the 
vertical thickness of the disc, the disc seen in projection at 
orbital phase 0.5 obscures the irradiated secondary from its pole down 
towards lower latitudes when $\cos i = (R_2/a) (1 + r_d/a)^{-1}$,
where $R_2$ is the radius of the secondary. For $q$ = 0.17 we find 
$R_2/a$ = 0.24 (Warner 1995), which gives $i = 80.5^{\circ}$. 
It happens, therefore, that in DD Cir
the inclination is such that a large fraction of one hemisphere of the
irradiated secondary is obscured by the optically thick accretion disc. With
the above parameters we find that the disc obscures the primary from its 
pole down to a latitude $l \sim 20^{\circ}$ (see Fig.~\ref{ddcirdr}), and that as a result 
$\sim$ 30\% of the surface area of
the secondary is obscured at orbital phase 0.5. Such obscuration is of 
course the case for all high inclination CVs, but in general the secondary
contributes such a small fraction of the total luminosity that the effect is 
not detectable. In DD Cir, however, we have a unique CV with high inclination
and substantial luminosity of one side of the secondary, so we may hope to
see secondary eclipses in the light curve.
    
\begin{figure}
\centerline{\hbox{\psfig{figure=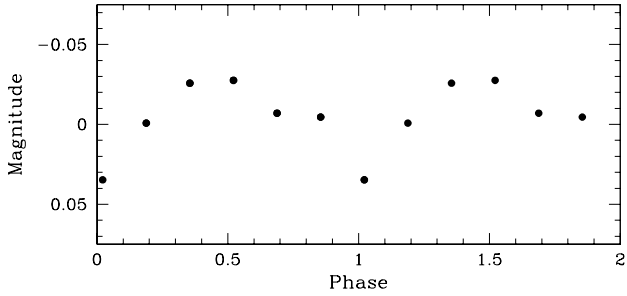,width=8.3cm}}}
  \caption{The average light curve of DD Cir (run S6421), folded on the
665 s periodicity. }
 \label{lcmeanddcir}
\end{figure}

DD Cir has one more property of interest. In runs S6416, S6421 and
S6425 the FTs show distinct periodic signals at 662 s, 665 s and at 673 s
respectively, all with amplitudes $\sim$ 0.025 mag. Note that for the analysis
of this $\sim$670 s signal in the various runs, we have removed the eclipses in each 
of the runs. The differences in period 
are within the uncertainties due to noise and data length. Furthermore, in 
S6399 there is a signal with amplitude 0.011 mag at 668 s, which is only at the
level of the noise, but supports the possible existence of a persistent
periodicity. In S6404 the amplitudes near 670 s are $<$ 0.01 mag, but a small
amount of noise in antiphase with the signal could cause cancellation. The
mean light curve for S6421, which has the clearest signal, is shown in 
Fig.~\ref{lcmeanddcir}, where the amplitude is 0.026 mag. The FT for the 
entire data set, excluding S6404, naturally has an alias problem; the 
two highest peaks are at 666.08 s and its one day alias 671.28 s, both with 
amplitudes of 0.017 mag.

The $\sim$670 s signal is indicative of the rotation period of the primary
or its orbital sideband. It would be valuable to extend observations of DD Cir,
with larger telescopes, before it fades further.

\subsection{TV Corvi}

    TV Crv is an SU UMa type dwarf nova with a quiescent magnitude V $\sim$ 19
and superoutbursts spaced about a year apart (Levy et al.~1990). During the
superoutburst in June 1994 superhumps were observed with a period $P_{sh}$ of
1.56 $\pm$ 0.02 h (Howell et al.~1996). Superhumps usually have periods a few
percent longer than the orbital periods (Warner 1995), so a $P_{orb}$ $\sim$ 
1.50 h is expected. No spectroscopic or photometric periods in quiescence have
hitherto been obtained.
   
We observed TV Crv at minimum light in the hope of detecting an orbital
modulation. Our observations are listed in Table 1. Having found a clear
orbital modulation on the first night, we observed on the second night for 
just sufficient time to include another orbital hump. The double humped profile
of the light curve puts a great deal of power into the first harmonic, which
has three aliases in the FT; we use these to find three possibilities for 
the fundamental period: 1.414, 1.461 and 1.512 h. In addition, there are three
aliases at the fundamental itself, at 1.506, 1.617 and 1.747 h. Knowing the
superhump period enables unambiguous selection of $P_{orb}$ = 1.509 h from this
suite. The ephemeris for times of maxima is given by

\begin{equation}
  {\rm HJD_{max}} = 2452372.3371 +  0{\fd}06288 (\pm 13) \, {\rm E}.
 \label{ephtvcrv}
\end{equation}

\begin{figure}
\centerline{\hbox{\psfig{figure=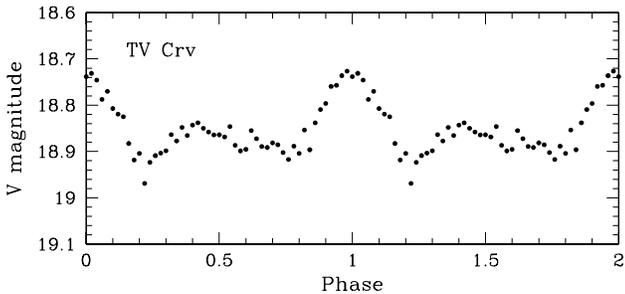,width=8.3cm}}}
  \caption{The average light curve of TV Crv at quiescence, folded on the orbital period 
of 1.509 h.}
 \label{lctvcrv}
\end{figure}

In Fig.~\ref{lctvcrv} we show the mean light curve, co-added at the orbital
period. The range of the orbital modulation is 0.2 mag. The superhump
excess  ($P_{sh} - P_{orb}$)/$P_{orb}$ = 0.034, and the beat period 
$P_b$ = ($P_{orb}^{-1}  - P_{sh}^{-1}$)$^{-1}$ = 1.9 d. 
These values are similar to those of other SU UMa stars with orbital
periods near 1.5 h (Warner 1995).

\subsection{V655 Coronae Austrinae}

\begin{figure}
\centerline{\hbox{\psfig{figure=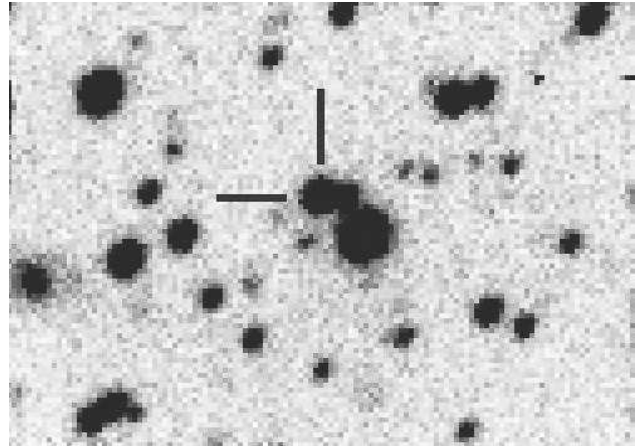,width=8.3cm}}}
  \caption{The finding chart of V655 CrA. The nova remnant is marked with bars. The field of view
is 50 by 34 arcsec, north is up and east is to the left.}
 \label{fcv655cra}
\end{figure}

       Nova Coronae Austrinae was discovered on objective prism plates in
June 1967 and was very poorly observed. Downes et al. (2001) give a
magnitude of 17.6 at quiescence, measured on a J plate. No observations of
V655 CrA at quiescence have previously been reported; our photometric
runs are listed in Table 1. A CCD frame of the vicinity of V655 CrA is
illustrated in Fig.~\ref{fcv655cra}; 
this shows that the nova remnant is a member of a
close triplet of stars. The star that we have found to be variable is 
marked.

\begin{figure}
\centerline{\hbox{\psfig{figure=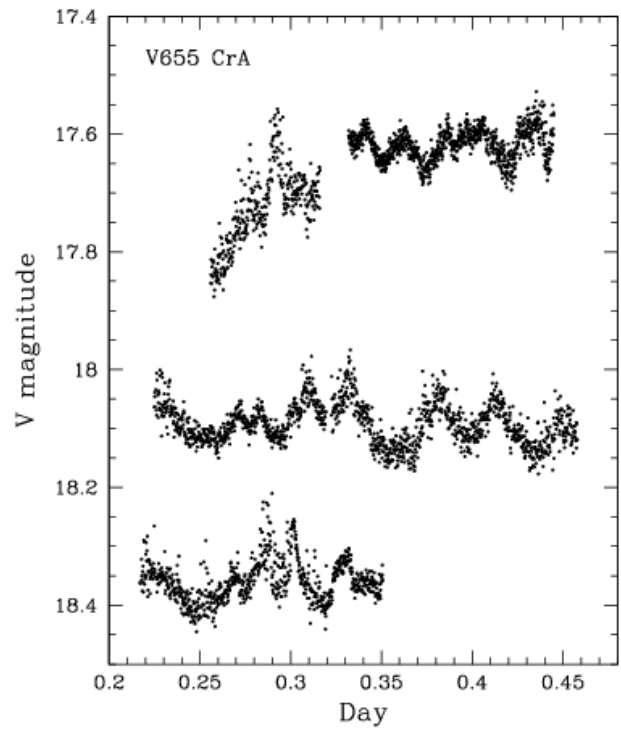,width=8.3cm}}}
  \caption{The light curves of V655 CrA, obtained in August/September 2002. The light curves of runs
S6512, S6517, and S6521 have been displaced vertically downwards by 0.25, 0.50, and 0.75 mag, respectively, 
for display purposes.}
 \label{lcv655cra}
\end{figure}

The light curves of V655 CrA are displayed in Fig.~\ref{lcv655cra}. There are
large variations around a mean magnitude of 17.6. Although there are clearly
preferred time scales associated with the variations, we find nothing in the
FTs indicative of repetitive modulations -- as we added more data, the FT
changed in character and in the positions of the principal peaks.

\subsection{CP Crucis}

     CP Cru was Nova Crucis 1996, discovered by Liller at V = 9.2 (Liller
1996) and observed at V = 19.48 in March 1998 (Downes \& Duerbeck
2000). It was a very fast nova ($t_2 \sim 4$ d) and is seen in an 
HST image to have a $0.6'' \times 0.6''$ shell (Downes \& Duerbeck 2000). 
True maximum light was probably missed.

    Our observations are listed in Table 1 and show that the nova was at 
V $\sim$ 19.6 in March 2002, and therefore is now probably near to its 
quiescent level. Such a fast nova, seen at the relatively high inclination 
that we deduce below, should have an eruption range $\sim$ 14.4 mag 
(Figure 5.4 of Warner 1995) and therefore could have reached 
V $\sim$ 5.2 at maximum. This is in agreement
with Downes \& Duerbeck, who deduce an absolute magnitude M$_{\rm V}$ $\sim$ -5.3
which is about 3.7 mag too faint for the speed of the nova.

\begin{figure}
\centerline{\hbox{\psfig{figure=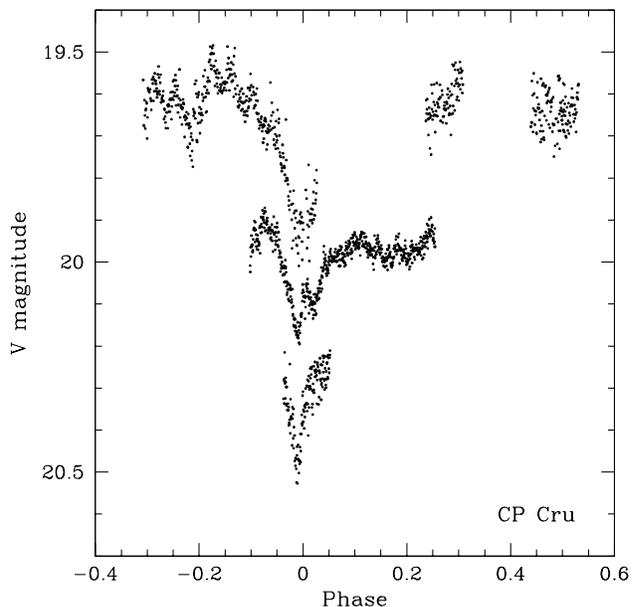,width=8.3cm}}}
  \caption{The light curves of CP Cru, taken in March 2002, folded on the orbital period of 22.7
h. The light curves of runs S6303 and S6304 have been displaced downwards by 0.3 and 0.6 mag, respectively, for
display purposes.}
 \label{lccpcru}
\end{figure}

   The light curves of CP Cru are shown in Fig.~\ref{lccpcru} and reveal 
that it has shallow eclipses, about 0.25 mag deep, giving it an 
inclination $\sim 70^{\circ}$. We
observed CP Cru just sufficiently to determine its orbital period
unambiguously (the initial observations gave several possible aliases; we
chose our subsequent observing times to eliminate or confirm these). The
orbital period is 22.7 h\footnote{The $P_{orb}$ reported earlier (Warner 2002) was 
mistakenly the first harmonic of the actual period.}, which is the third longest period 
known for a classical nova. The secondary in CP Cru must be considerably evolved.
The ephemeris for mid-eclipse is

\begin{equation}
  {\rm HJD_{min}} = 2452346.620 +  0{\fd}0944 (\pm 1) \, {\rm E}.
 \label{ephcpcru}
\end{equation}

\subsection{V794 Ophiuchi}

V794 Oph was discovered by Burwell \& Hoffleit (1943) on an objective prism
plate taken in July 1939 and became a slow nova with $t_3$ = 220 d, resembling HR Del
in its light curve (Duerbeck 1987). Szkody (1994) gave V = 17.70 and (B--V) = 1.5, measured
in 1989, and Hoard et al. (2002) give J, H, and K magnitudes of the same star, identified
on Duerbeck's (1987) finding chart. Ringwald, Naylor \& Mukai (1996) obtained a spectrum
of the star, showing a featureless continuum increasing strongly to long wavelengths.

At maximum, V794 Oph was $m_{pg}$ = 11.7. The range of such a slow nova would be expected
to be 8--11 mag, depending on its orbital inclination (Figure 5.4 of Warner 1995). The candidate
star at V $\sim$ 17.7 would therefore seem rather too bright to be the nova remnant, and this 
conclusion is supported by the spectrum.

We have obtained high speed photometry of the candidate, see Table 1, and find no variability
in it. We have not found rapid variability of any other stars in the field,
either of the candidate, or of stars near the position given by Burwell \& Hoffleit, which differs
from Duerbeck's candidate by nearly 1$'$. It is not clear by how much maximum brightness was missed,
but we suggest that the true nova remnant is likely to be considerably fainter than the
candidate's 17.7. Our magnitude estimate for the V794 Oph candidate is somewhat uncertain due to
its red colours.

\subsection{V992 Scorpii}

\begin{figure*}
\centerline{\hbox{\psfig{figure=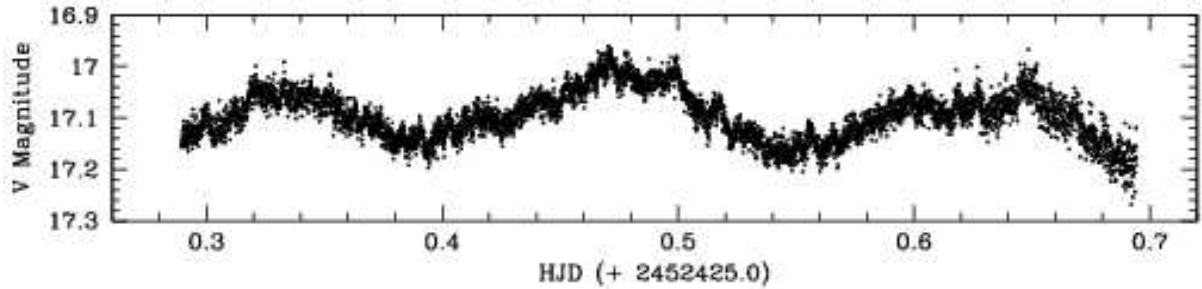,width=16.0cm}}}
  \caption{The light curve of V992 Sco (run S6398), taken on 2002 May 30. }
 \label{lcv992sco}
\end{figure*}

Nova Scorpii 1992 was discovered at V $\sim$ 8.2 by Camilleri (1992) and
brightened to V = 7.26 four days after discovery (Kilmartin 1992). It was a
slow nova, the light curve of which is shown by Smith et al. (1995).
    
\begin{figure}
\centerline{\hbox{\psfig{figure=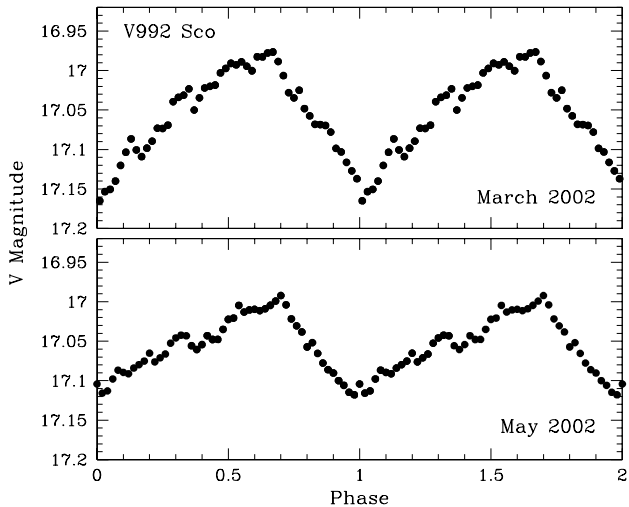,width=8.3cm}}}
  \caption{The mean light curve of V992 Sco in March 2002 (upper panel) and May 2002 (lower panel),
folded on the 3.686 h period.}
 \label{lcv992scoav}
\end{figure}

Our photometric observations of V992 Sco are listed in Table 1 and show
the nova to have faded to V $\sim$ 17.1 by early 2002, giving an eruption range
of at least 9.8 mag. The light curve of the longest run in the May 2002 set of runs (S6398)
is shown in Fig.~\ref{lcv992sco}. An FT of this data set, followed by a non-linear least squares
fit to the fundamental and first harmonic, delivers a fundamental period of
3.686 h with an amplitude of 0.044 mag and a first harmonic with an
amplitude of 0.018 mag. Similar treatment of the March 2002 runs gives the
independent estimates 3.685 h at 0.070 mag and the harmonic at 0.015 mag.
Fig.~\ref{lcv992scoav} shows the mean light curves of the March and May runs, co-added at 
the 3.686 h period, which we interpret as $P_{orb}$ for V992 Sco. 
The low amplitude of the orbital modulation suggests an intermediate orbital inclination,
probably $\sim$40$^{\circ}$.
    
The FTs show no other coherent periodicities in the light curves.

\subsection{EU Scuti}

       EU Sct was Nova Scuti 1949, discovered on 31 July of that year and
reaching a photographic magnitude of 8.4 at maximum brightness 5 days
later. It was a moderately fast nova with $t_3$ = 42 d. Szkody (1994) measured
V = 17.37 and (B--V) = 2.66 in 1988 for the candidate star indicated in
Deurbeck's (1987) finding chart. Szkody pointed out the extreme redness of
this star, as also did Weight et al (1994), who proposed  (by comparison 
with the recurrent novae RS Oph and T CrB which have red giant companions)
that EU Sct might be a recurrent nova. There has been no detailed
spectroscopic study of EU Sct in quiescence.
    
Our observations of EU Sct are listed in Table 1. The photometric calibration
that we normally apply is not appropriate for an object with such red colours. We 
therefore do not list the mean magnitude of EU Sct in Table 1. We find no rapid
variation in the candidate star -- which, being very red, appears much
brighter on our CCD frames than on the photographic finding chart and is
easy to measure accurately. However, there is a 19.1 mag companion $4''$
south of the candidate, which is difficult to observe but in the first of 
our runs has possible photometric variations. If this is the true candidate 
then the eruption range was 10.7 mag, which is compatible with a nova 
having $t_3 = 42$ d if it has low orbital inclination 
(Fig. 5.4 of Warner 1995). Spectra of the
two objects are required to test our suspicions.

\subsection{V373 Scuti}

   V373 Sct was discovered as a nova on 15 Jun 1975, but it was
subsequently found that maximum light, at V = 7.1, had occurred just over a
month earlier (Duerbeck 1987). It was a moderately fast nova with $t_3 = 85$ d.
The spectrum obtained by Ringwald et al. (1996) in 1991
showed strong Balmer emission on a blue continuum, and very strong HeII
4686\AA and the CIII/NIII blend at 4650\AA, which is often indicative of a
magnetic system. Their photometry gave a V magnitude of 18.7.

\begin{figure}
\centerline{\hbox{\psfig{figure=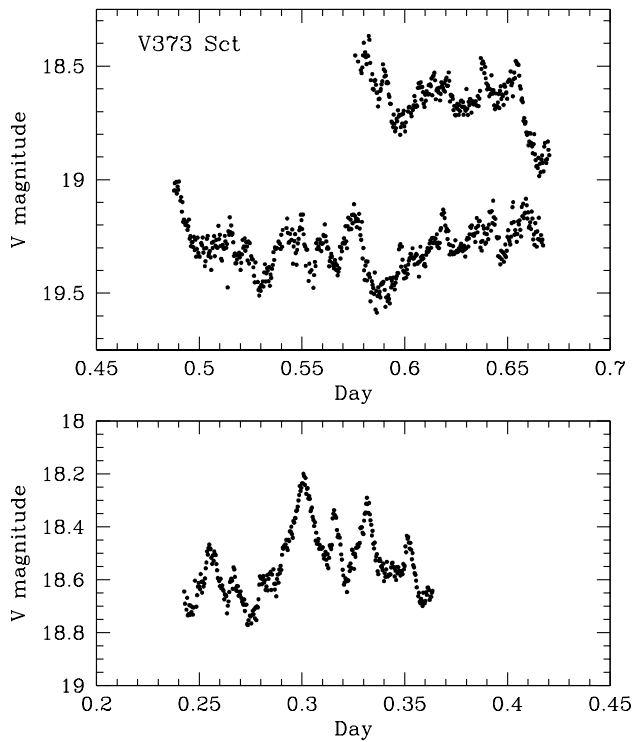,width=8.3cm}}}
  \caption{The light curves of V373 Sct, obtained in April 2002 (upper panel) and
August 2002 (lower panel). The light curve of run S6361 (upper panel) has been displaced 
downwards by 0.7 mag for display purposes.}
 \label{lcv373sct}
\end{figure}

   Our observations are listed in Table 1. We also find a mean magnitude
near 18.7, showing that V373 Sct has settled down at minimum light. The
eruption range of 11.6 mag and value of $t_3$ combine to suggest a high
inclination orbit. However, the light curves, seen in Fig.~\ref{lcv373sct}, 
show great activity but no hint of any orbital modulation.
      
\begin{figure}
\centerline{\hbox{\psfig{figure=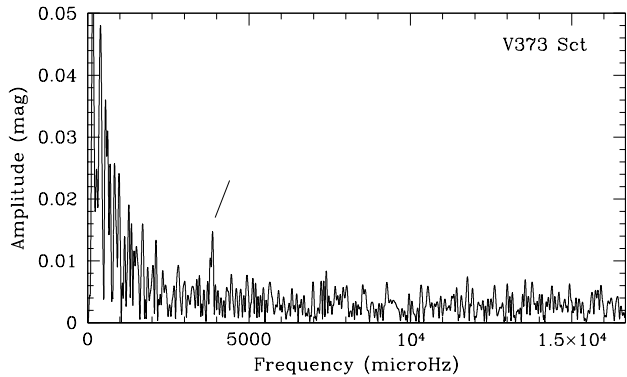,width=8.3cm}}}
  \caption{The Fourier transform of V373 Sct (run S6361). The peak at 258.3 s is marked by the
diagonal bar.}
 \label{ftv373sct}
\end{figure}

The large amplitude flickering, as with the strong ionic emission lines, is
also often a signature of a magnetic system. In V373 Sct we have found that
a third magnetic diagnostic is present -- a coherent luminosity modulation of
the kind seen in non-synchronous rotators such as DQ Herculis stars and
intermediate polars. The only significant feature in the FTs of the three 
light curves is a narrow spike at a period of 258.3 s in run S6361 -- 
see Fig.~\ref{ftv373sct}.
There is no significant power at this period in the FTs of the other two 
runs -- but there is considerable noise in this region due to the flickering. 
There is no significant amplitude at the first harmonic of the signal.
   
\begin{figure}
\centerline{\hbox{\psfig{figure=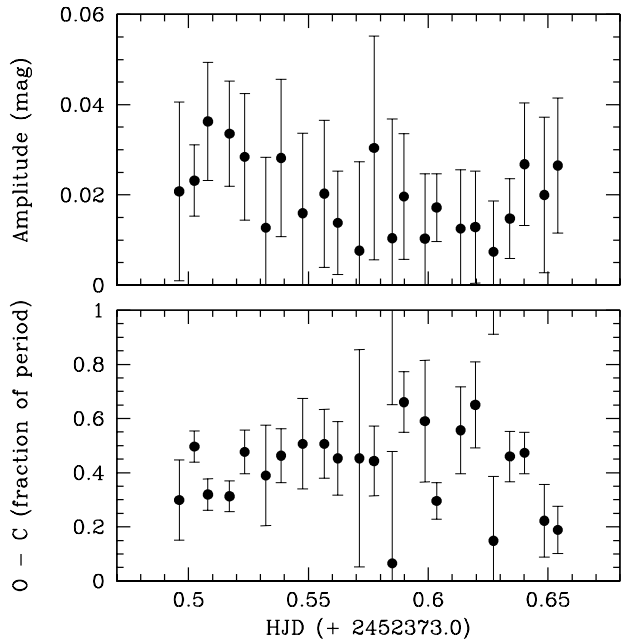,width=8.3cm}}}
  \caption{An O--C diagram (lower panel) of V373 Sct (run S6361) at 258.3 s. 
The upper panel shows the amplitude of the 258.3 s throughout the run.}
 \label{v373sctomc}
\end{figure}

Fig.~\ref{v373sctomc} shows the O--C diagram and amplitude plot for run S6361.
Five cycles of the signal, with 50\% overlap, have been fitted to a sinusoid
with a period of 258.3 s. There is an increase in the length of the error 
bars on the amplitude points where steep slopes occur in the light curve. It is
difficult to filter the large amplitude rapid variations out of the light 
curve without removing part of the coherent signal, which is of similar time 
scale, so we have merely analysed the raw light curve. The O--C phase 
variations are consistent with a signal of constant period, the mean 
amplitude of which is $\sim$ 0.02 mag.
    
A modulation at 258 s is considerably longer than any DNOs observed in
CVs (which usually have periods much less than 100 s). However, the three
old novae V533 Her, DQ Her and GK Per show (or have shown, in the case
of V533 Her, where the signal is no longer visible) highly stable
modulations at 63.63, 71.07 and 351.34 s respectively (see, e.g., the review
in Warner 1995). 
If the 258 s periodicity proves to be persistent and coherent then V373 Sct
will join this group of rapid rotators.

\section{Summary}

       The most important result from our survey is that we have detected
eclipses in three old novae: BY Cir with $P_{orb}$ = 6.76 h, DD Cir with $P_{orb}$ =
2.339 h (placing it squarely in the ``orbital period gap''), and CP Cru with 
$P_{orb}$ = 22.7 h, which is the third longest period known for a classical nova. We
also find a fourth orbital period, from the presence of a modulation at 
3.686 h in the old nova V992 Sco. In addition, DD Cir has a $\sim$ 670 s low amplitude
photometric modulation and the old nova V373 Sct has a 258.3 s
modulation; these are signatures of magnetic primaries with these as their
spin period, or orbital side bands. There is some evidence that DD Cir has a
0.3 mag reflection effect that is partly obscured at phase 0.5, producing a
secondary eclipse caused by the optically thick accretion disc.

    The old novae V842 Cen and V655 CrA are highly active photometrically
but show no clear periodic modulation, indicating that they are low
inclination systems.

   We have also measured $P_{orb}$ = 1.509 h for the SU UMa star TV Crv from
an orbital modulation seen during quiescence. Only the superhump period,
observed during superoutburst, was previously known for this star.

   Our observations of V794 Oph and EU Sct throw doubt on the currently
proposed identification of these old novae.

\section*{Acknowledgments}
We thank Dr. D.~O'Donoghue for the use of his EAGLE program for 
Fourier analysis of the light curves. 
PAW is supported by funds made available from the National Research
Foundation and by strategic funds made available to BW from the
University of Cape Town. BW's research is supported by the University.

\end{document}